\documentclass[journal]{IEEEtran}
\usepackage{amsmath,latexsym,amssymb,stmaryrd,pifont}
\usepackage{setspace,stackrel,tikz,graphicx}
\usepackage{exscale,relsize,subfig,textcomp,stackrel,setspace,float}
\usepackage{mdframed,pifont}
\usepackage{multicol,xfrac}
\usepackage{multirow, booktabs}
\usepackage{mathtools}
\usepackage{tabularx}
\usepackage[input-decimal-markers=.]{siunitx}
\usepackage{longtable}
\usepackage[T1]{fontenc}
\usepackage[font=footnotesize]{caption}
\usepackage{lipsum}
\usepackage{multicol}
\usepackage{graphicx}
\usepackage[utf8]{inputenc}
\pdfoutput=1
\usepackage{amsthm}
\usepackage{mathtools,amssymb,lipsum}

\usepackage{cuted}

\begin{document}

\title{\vspace{0cm}\LARGE Dual-Mode Time Domain Multiplexed Chirp Spread Spectrum\vspace{0em}}
\makeatletter
\patchcmd{\@maketitle}
  {\addvspace{0\baselineskip}\egroup}
  {\addvspace{0\baselineskip}\egroup}
  {}
  {}
\makeatother
\author{Ali Waqar Azim, Ahmad Bazzi, Mahrukh Fatima, Raed Shubair, Marwa Chafii
\thanks{Ali Waqar Azim is with Department of Telecommunication Engineering,  University of Engineering and Technology,  Taxila,  Pakistan (email: aliwaqarazim@gmail.com).}
\thanks{Ahmad Bazzi and Raed Shubair is with with Engineering Division, New York University (NYU) Abu Dhabi, 129188, UAE (email: \{ahmad.bazzi,raed.shubair\}@nyu.edu).}
\thanks{Mahrukh Fatima is with National Defence University,  Islamabad,  Pakistan (email: mahrukh@ndu.edu.pk).}
\thanks{Marwa Chafii is with Engineering Division, New York University (NYU) Abu Dhabi, 129188, UAE and NYU WIRELESS, NYU Tandon School of Engineering, Brooklyn, 11201, NY, USA (email: marwa.chafii@nyu.edu).}}
\maketitle
\begin{abstract}
We propose a dual-mode (DM) time domain multiplexed (TDM) chirp spread spectrum (CSS) modulation for spectral and energy-efficient low-power wide-area networks (LPWANs). DM-CSS modulation that uses both the even and odd cyclic time shifts has been proposed for LPWANs to achieve noteworthy performance improvement over classical counterparts. However, its spectral efficiency (SE) is half of the in-phase and quadrature (IQ)-TDM-CSS scheme that employs IQ components with both up and down chirps, resulting in a SE that is four times relative to Long Range (LoRa) modulation. Nevertheless, the IQ-TDM-CSS scheme only allows coherent detection. Furthermore, it is also sensitive to carrier frequency and phase offsets, making it less practical for low-cost battery-powered LPWANs for Internet-of-Things (IoT) applications. DM-CSS uses either an up-chirp or a down-chirp. DM-TDM-CSS consists of two chirped symbols that are multiplexed in the time domain. One of these symbols consisting of even and odd frequency shifts (FSs) is chirped using an up-chirp. The second chirped symbol also consists of even and odd FSs, but they are chirped using a down-chirp. It shall be demonstrated that DM-TDM-CSS attains a maximum achievable SE close to IQ-TDM-CSS while also allowing both coherent and non-coherent detection. Additionally, unlike IQ-TDM-CSS, DM-TDM-CSS is robust against carrier frequency and phase offsets. 
\end{abstract}
\IEEEpeerreviewmaketitle
\vspace{-3mm}
\section{Introduction}
\IEEEPARstart{T}{he} rudimentary idea for Internet-of-Thing (IoT) applications is to communicate between battery-powered devices/sensors by consuming low power to extend the battery lifetime of the terminals. In this regard, low-power wide-area networks (LPWANs) are of utmost significance. One of the emerging technologies of LPWANs is the Long Range (LoRa) wide-area network (LoRaWAN), which uses LoRa as the physical layer modulation scheme. 

LoRa is a proprietary derivative of chirp spread spectrum (CSS) modulation, developed by the Semtech corporation, capable of trading off sensitivity with data rates for fixed channel bandwidths \cite{lorawan_spec,lora_mod_basics}. Even though the Semtech corporation never published the details of LoRa modulation, Vangelista in \cite{lora} has provided a comprehensive theoretical explanation with an optimal low-complexity detection process based on discrete Fourier transform (DFT). The scalable parameter of the LoRa modulation is the \textit{spreading factor}, \(\lambda\), where \(\lambda = \llbracket 6,12\rrbracket\). \(\lambda\) is, in fact, equal to the number of bits that a LoRa symbol can transmit. Moreover, LoRa symbols are defined using \(M\) cyclic time shifts of the chirp, which are the information-bearing elements. These cyclic time shifts correspond to frequency shifts (FSs) of the complex conjugate of the chirp signal, i.e., down-chirp signal; thus, LoRa can be regarded as FS chirp modulation, where \(\lambda = \log_2(M)\).

Besides the broad adoption and numerous benefits of LoRa, one of the limiting factors is that it achieves low achievable rates in all three bands it utilizes. Several recent studies proposed spectral-efficient CSS modulation schemes as possible alternatives to LoRa. The waveform design of these CSS alternatives is comprehensively elucidated in \cite{azim2022survey}. It is noteworthy that these CSS variants can have different properties, e.g., some possess constant envelope properties and are single chirp, whereas others use multiple chirps in the symbol structure and do not retain constant envelope properties. Possessing a constant envelope is desirable; however, the schemes possessing a constant envelope generally have low spectral efficiencies, which could be an influential limiting factor. 

Another aspect to consider for the CSS alternatives to LoRa is that they can achieve different maximum achievable spectral efficiencies. Moreover, most LoRa alternatives aim to improve spectral efficiency (SE), energy efficiency (EE), or both. Some of the most promising recently proposed alternatives to LoRa are in-phase and quadrature (IQ)-CSS \cite{iqcss}, slope-shift keying interleaved chirp spreading LoRa (SSK-ICS-LoRa) \cite{ssk_ics_lora}, dual-mode CSS \cite{dm_css_ieee}, and the time domain multiplexed (TDM) schemes \cite{tdm_lora}, such as TDM-CSS and in-phase and quadrature (IQ)-TDM-CSS. It is accentuated that here we only mention a subset of energy-efficient CSS modulations available in the literature. Numerous other CSS schemes exist in the state-of-the-art; interested readers are referred to \cite{azim2022survey} for more details. IQ-CSS is another multiple chirp modulation that encodes information bits on both in-phase and quadrature components of the chirp signal. SSK-ICS-LoRa uses up chirps, down chirps, interleaved up-chirps, and interleaved down chirps to expand the symbol set and hence can carry two additional bits per symbol relative to LoRa. DM-CSS  simultaneously multiplexes even and odd chirp symbols, use phase shifts (PSs) of \(0\) and \(\pi\) radians for these even and odd chirp symbols, and employs either up-chirp or down-chirp signal. In TDM-CSS, two chirped symbols are multiplexed in the time domain, each having a different chirp slope, i.e., one (chirped) symbol is attained using an up-chirp, whereas the other one is attained using a down-chirp. The fundamental idea of IQ-TDM-CSS is similar to that of TDM-CSS; however, unlike TDM-CSS, IQ-TDM-CSS uses both the IQ components of the un-chirped symbols. 
It may be noticed that the SE of the DM-CSS and TDM schemes is higher than that of SSK-ICS-LoRa and classical LoRa. Note that if LoRa transmits \(\lambda\) bits per symbol, then SSK-ICS-LoRa, IQ-CSS, TDM-CSS, DM-CSS, and IQ-TDM, respectively, transmit \(\lambda+2\), \(2\lambda\), \(2\lambda\), \(2\lambda+1\) and \(4\lambda\) bits per symbol of the same duration. 

Nevertheless, all these schemes also have some noteworthy shortcomings.IQ-CSS is very sensitive to the carrier frequency offset, and its maximum achievable SE is less than that of DM-CSS and IQ-TDM-CSS. Besides being capable of both coherent and non-coherent detection and providing improved EE, SSK-ICS-LoRa does not significantly improve the SE relative to LoRa and its other counterparts. TDM-CSS symbols can also be detected coherently and non-coherently; however, its maximum achievable SE is lesser than DM-CSS and IQ-TDM-CSS. Besides being sensitive to carrier frequency offset, for IQ-TDM-CSS, only highly complex coherent detection is possible. DM-CSS offers improved EE relative to LoRa and allows both coherent and non-coherent detection; however, the use of PSs allows the non-coherent detection to be only feasible if the channel phase rotation is less than \(\sfrac{\pi}{2}\), making it less practical. Furthermore, its maximum achievable SE is less than that of IQ-TDM-CSS.

In this work, we propose the DM-TDM-CSS scheme that can achieve almost the same maximum achievable SE as IQ-TDM-CSS and eliminate its shortcomings. In other words, both coherent and non-coherent detection can be applied for DM-TDM-CSS and is more robust against carrier frequency and phase offsets. In DM-TDM-CSS, we amalgamate the precepts of a modified version of DM-CSS and TDM-CSS. To be more precise, like DM-CSS, we use the even and the odd FSs; however, unlike DM-CSS, we do not use any PSs for the even and the odd FSs, allowing more practical coherent detection. Moreover, like TDM-CSS, two chirped symbols are multiplexed in the time domain. Both un-chirped symbols have unique even and odd FSs; however, one is chirped using an up-chirp and the other with a down-chirp. Unlike DM-CSS, which uses either up-chirp or down-chirp symbols for chirping the un-chirped symbol, DM-TDM-CSS uses both simultaneously. The symbols structure of DM-TDM-CSS allows for achieving a maximum SE, which is only \(4\) bits less than that of IQ-TDM-CSS. 

The contributions of this work can be summarized as follows:
\begin{enumerate}
\item We propose the DM-TDM-CSS scheme as an alternative to state-of-the-art CSS schemes (including LoRa). DM-TDM-CSS inherits the advantageous properties of both DM-CSS and TDM-CSS while avoiding the limitations of both schemes. Thus, the resulting scheme is not only energy and spectral-efficient but is also robust against the carrier frequency and phase offsets. 
\item We comprehensively explain the transceiver design of the proposed DM-TDM-CSS. The waveform generation and the coherent and non-coherent detection mechanisms are elaborately presented.
\item We mathematically determine if the DM-TDM-CSS symbols are orthogonal to each other or not. It shall be demonstrated that the even and odd FSs in the up-chirp symbol cause interference with the even and odd FSs in the down-chirp symbol and vice versa. This interference among the time domain multiplexed symbols results in non-orthogonal DM-TDM-CSS symbols. 
\item Through mathematical analysis, we estimate the interference caused by the two TDM chirped symbols at the receiver. The results shall affirm the conclusions of the orthogonality analysis that the two TDM symbols cause interference among each other. 
\item We evaluate the performance of DM-TDM-CSS considering different performance metrics: (i) SE versus required signal-to-noise ratio (SNR) per bit for a target bit error rate (BER);  (ii) BER performance in an additive white Gaussian noise (AWGN) and a fading channel; and (iii) BER performance considering phase and frequency offsets.
\item We provide closed-form expressions on the interference terms on both the up/down chirped symbols. We compute the signal-to-interference ratio (SIR) expressions thanks to the closed-form expressions. We show that the interference vanishes by increasing \(\lambda\).
\end{enumerate}
\vspace{-4mm}
\section{Preliminaries}
\vspace{-2mm}
\subsection{Basic Definitions}
In this section, we have provided brief definitions of these parameters for the clarity of the readers.
\subsubsection{Bandwidth}
Bandwidth, \(B\) is the range of frequencies in Hertz (Hz) occupied by a CSS symbol. \(B\) is divided into \(M\) frequencies, where the separation between two adjacent frequencies is \(\Delta f\) Hz; therefore, \(B = M\Delta f\) Hz.
\subsubsection{Symbol Period}
Symbol period, \(T_\mathrm{s}\) is the time in which one CSS symbol, occupying bandwidth, \(B\), can be transmitted. \(T_\mathrm{s}\) is linked to \(\Delta f\) as \(\Delta f = \sfrac{1}{T_\mathrm{s}}\). 
\subsubsection{Bit Rate}
Bit rate \(R\) in bits/s is the number of bits that can be transmitted in \(T_\mathrm{s}\). 
\subsubsection{Spectral Efficiency}
It is the achievable \(R\) per \(B\) and is given as \(\eta = \sfrac{R}{B}\). 
\subsubsection{Energy Efficiency}
It is the required SNR for correct bit detection at a given BER.
\vspace{-5mm}
\subsection{System Model}
Without loss of generality, we consider a chirped symbol in CSS modulation composed of two components: (i) an un-chirped symbol and (ii) a spreading symbol that spreads the information in the bandwidth, \(B\). The un-chirped symbol is a pure sinusoid when only one FS \(k\) is activated, or it can be a combination of multiple sinusoids in case multiple FSs are used. When the un-chirped symbol is spread, the FS(s) have an injective mapping to cyclic time-shift(s). Moreover, the spreading symbol can have different slope rates \cite{ssk_lora, dcrk_css}. 

We consider that the occupied bandwidth is \(B = \sfrac{M}{T_\mathrm{s}}\) that corresponds to the availability of \(M\) FSs implying that \(k \in \left[0, M-1\right]\).  In the discrete time, we denote the CSS (chirped) symbol consisting of \(M\) samples by \(s(n) = g(n)c_\gamma(n)\), for \(n=\llbracket 0, M-1\rrbracket\), where \(g(n) \) is the un-chirped symbol, and \(c_\gamma(n)\) is the spreading symbol given as \(c_\gamma(n) = \exp\left\{j\frac{\pi}{M}\gamma n^2\right\}\), where \(j^2 = -1\) and \(\gamma\) is the slope rate.  Based on the type of CSS modulation, \(g(n)\) can have different symbol structures. Moreover,  when \(\gamma = 1\), the spreading symbol corresponds to up-chirp, i.e., \(c_\mathrm{u}(n)= \exp\left\{j\frac{\pi}{M} n^2\right\}\). Conversely, if \(\gamma = -1\), the spreading symbol is a down-chirp denoted as \(c_\mathrm{d}(n)= \exp\left\{-j\frac{\pi}{M} n^2\right\}\). Typically, an up-chirp symbol is used to spread the information in most CSS modulations. 

The discrete-time baseband received symbol is:
\vspace{-2mm}
\begin{equation}\label{eq1}
y(n) = hs(n) + w(n),
\end{equation}
where \(h\) is the complex channel gain, and \(w(n)\) are the AWGN samples having single-sided  noise power spectral density of \(N_0\), and noise variance of \(\sigma_n^2=N_0B\). It may be noticed that in LPWANs, CSS symbols maintain a narrow bandwidth of \(500\) kHz or smaller. Therefore, a flat fading channel can have a constant attenuation over the entire \(B\). Thus, it can be considered equal to unity if channel state information (CSI) is known in the simplest of cases.
\vspace{-5mm}
\section{Dual-Mode Time Domain Multiplexed Chirp Spread Spectrum}
\begin{figure}[tb]\centering
\includegraphics[trim={0 0 0 0},clip,scale=0.98]{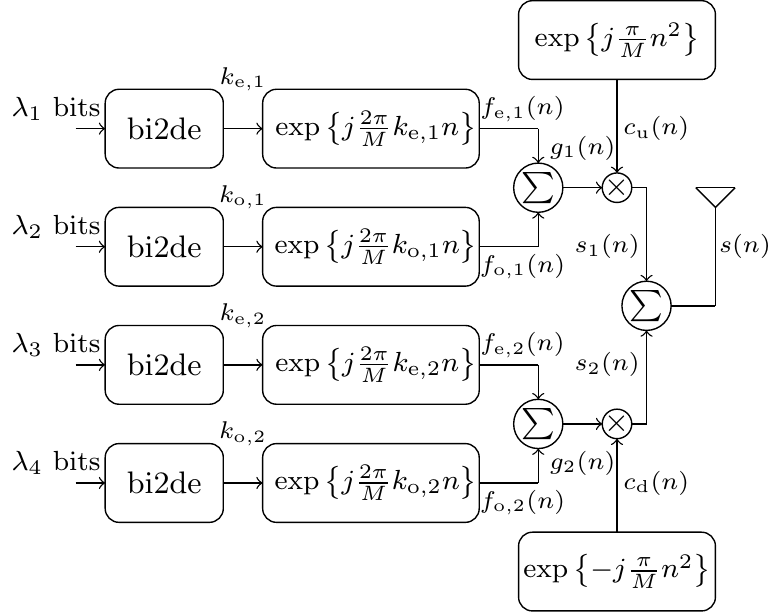}
  \caption{DM-TDM-CSS transmitter architecture. }
\label{fig1}
\end{figure}
\vspace{-3mm}
\subsection{Transmission}
The transmitter architecture of DM-TDM-CSS is provided in Fig. \ref{fig1}. In DM-TDM-CSS, two chirped symbols are multiplexed in the time domain; therefore, two different un-chirped symbols are needed. For each un-chirped symbol, one even and one odd frequency is activated. Consider that for each un-chirped symbol, \(M\) frequencies are available. Among these \(M\) frequencies, \(\sfrac{M}{2}\) frequencies are even and \(\sfrac{M}{2}\) frequencies are odd. The even activated frequencies for these un-chirped symbols are \(k_{\mathrm{e},1}\) and \(k_{\mathrm{e},2}\), whereas the odd activated frequencies are \(k_{\mathrm{o},1}\) and \(k_{\mathrm{o},2}\). Note that the even and odd frequencies are identified by indexes \(\tilde{k}_\mathrm{e}= 2k_\mathrm{e}\) and \(\tilde{k}_\mathrm{o}=2k_\mathrm{o}+1\), where \( k_\mathrm{e}\in \left[0,\sfrac{M}{2}-1\right]\) and \( k_\mathrm{o}\in \left[0,\sfrac{M}{2}-1\right]\). \(k_{\mathrm{e},1}\) and \(k_{\mathrm{o},1}\) are determined after binary-to-decimal (bi2de) conversion of bit sequences of lengths \(\lambda_1=\lambda-1\) and \(\lambda_2=\lambda-1\), respectively. On the other hand, \(k_{\mathrm{e},2}\) and \(k_{\mathrm{o},2}\) after bi2de conversion of bit sequences having lengths \(\lambda_3=\lambda-1\) and \(\lambda_4=\lambda-1\), respectively.

The first un-chirped symbol, \(g_1(n)\), is composed of two sinusoids. The first sinusoid, \(f_{\mathrm{e},1}(n)\), has an even activated frequency, \(k_{\mathrm{e},1}\), whereas the second sinusoid, \(f_{\mathrm{o},1}(n)\), has an odd activated frequency, \(k_{\mathrm{o},1}\). Then, \(g_1(n)\) is given as:
\begin{equation}\label{eq2}
\begin{split}
g_1(n) &= f_{\mathrm{e},1}(n) + f_{\mathrm{o},1}(n)\\
&= \exp\left\{j\frac{2\pi}{M} k_{\mathrm{e},1} n\right\} + \exp\left\{j\frac{2\pi}{M} k_{\mathrm{o},1} n\right\}.
\end{split}
\end{equation}

Similarly, the second un-chirped symbol, \(g_2(n)\), also consists of even frequency, \(k_{\mathrm{e},2}\), and odd frequency, \(k_{\mathrm{o},2}\), activated sinusoids, \(f_{\mathrm{e},2}(n)\), and \(f_{\mathrm{o},2}(n)\). \(g_2(n)\) is given as:
\begin{equation}\label{eq3}
\begin{split}
g_2(n) &= f_{\mathrm{e},2}(n) + f_{\mathrm{o},2}(n)\\
&= \exp\left\{j\frac{2\pi}{M} k_{\mathrm{e},2} n\right\} + \exp\left\{j\frac{2\pi}{M} k_{\mathrm{o},2} n\right\}.
\end{split}
\end{equation}

The next step is to spread the un-chirped symbols, \(g_1(n)\), and \(g_2(n)\). \(g_1(n)\) is then spread using an up-chirp, \(c_\mathrm{u}(n)\), whereas \(g_2(n)\) is spread using a down-chirp, \(c_\mathrm{d}(n)\) resulting in \(s_1(n)\) and \(s_2(n)\), i.e.,
\begin{equation}\label{eq4}
\begin{split}
s_1(n) &= g_1(n)c_\mathrm{u}(n)\\
&= \exp\left\{\!j\frac{\pi}{M}\left(2k_{\mathrm{e},1}n+n^2\right) \!\right\}\! +\! \exp\left\{\!j\frac{\pi}{M} \left(2k_{\mathrm{o},1}n+n^2\right)\!\right\},
\end{split}
\end{equation}
and
\begin{equation}\label{eq5}
\begin{split}
s_2(n) &= g_2(n)c_\mathrm{d}(n)\\
&= \exp\left\{\!j\frac{\pi}{M}\left(2k_{\mathrm{e},2}n-n^2\right) \!\right\}\! +\! \exp\left\{\!j\frac{\pi}{M} \left(2k_{\mathrm{o},2}n-n^2\right)\!\right\}.
\end{split}
\end{equation}

Afterward, these two chirped symbols, \(s_1(n)\) and \(s_2(n)\), are multiplexed in the time domain resulting in \(s(n)\), which is given as:
\begin{equation}\label{eq6}
\begin{split}
s(n) &= s_1(n)+s_2(n)\\
&= \exp\left\{j\frac{\pi}{M}\left(2k_{\mathrm{e},1}n\!+\!n^2\right) \right\}\! +\! \exp\left\{j\frac{\pi}{M} \left(2k_{\mathrm{o},1}n\! +\!n^2\right)\right\}\\
&+ \exp\left\{j\frac{\pi}{M}\left(2k_{\mathrm{e},2}n\!-\!n^2\right) \right\}\! +\! \exp\left\{j\frac{\pi}{M} \left(2k_{\mathrm{o},2}n\! -\!n^2\right)\right\}.
\end{split}
\end{equation}

The symbol energy of the DM-TDM-CSS symbol is \(E_\mathrm{s} = \sfrac{1}{M}\sum_{n=0}^{M-1}\vert s(n) \vert^2\). 
\vspace{-5mm}
\subsection{Detection}
This section presents coherent and non-coherent detection mechanisms for DM-TDM-CSS received symbols, \(y(n)\).  For clarity of exposition, we consider the following vectorial representations, \(\boldsymbol{y}= \left[y(0),y(1),\cdots,y(M-1)\right]^\mathrm{T}\), and \(\boldsymbol{s}= \left[s(0),s(1),\cdots,s(M-1)\right]^\mathrm{T}\), where \([\cdot]^\mathrm{T}\) is the transpose operator. 
\subsubsection{Coherent Detection}
\begin{figure}[h]\centering
\includegraphics[trim={0 0 0 0},clip,scale=1]{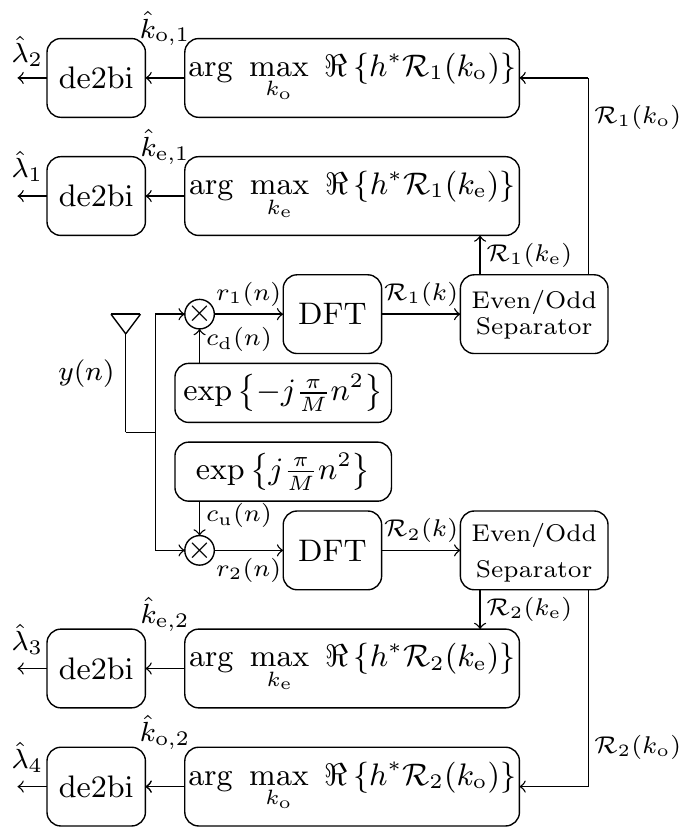}
  \caption{DM-TDM-CSS coherent detector architecture. }
\label{fig2}
\end{figure}
The coherent detector achitecture for DM-TDM-CSS is illustrated in Fig. \ref{fig2}. The coherent detection of DM-TDM-CSS symbols involves the estimation of the FSs of the un-chirped symbols, \(k_{\mathrm{e},1}\), \(k_{\mathrm{e},2}\), \(k_{\mathrm{o},1}\), and \(k_{\mathrm{o},2}\). Assuming that \(h\) is known at the receiver and the transmit symbols are equiprobable, the coherent detection dictates to maximize the probability of receiving \(\boldsymbol{y}\) when \(\boldsymbol{s}\) was sent given \(h\), i.e., \(\mathrm{prob}\left(\boldsymbol{y}\vert \boldsymbol{s},h\right)\). The likelihood function, \(\mathrm{prob}\left(\boldsymbol{y}\vert \boldsymbol{s},h\right)\) is given as:
\begin{equation}\label{eq7}
\begin{split}
\mathrm{prob}\left(\boldsymbol{y}\vert \boldsymbol{s},h\right) &= \left(\frac{1}{2\pi \sigma_n^2}\right)^M\exp\left\{\frac{\Vert\boldsymbol{y}-h\boldsymbol{s}\Vert^2}{2\sigma_n^2}\right\}\\
& = \rho \exp\left\{\frac{\Re\left\{\langle \boldsymbol{y},h\boldsymbol{s}\rangle\right\}}{\sigma_n^2}\right\},
\end{split}
\end{equation}
where \(\Vert\cdot\Vert^2\) evaluates Euclidean norm, \(\Re\{\cdot\}\) determines the real component of a complex-valued argument, and \[\rho= \left(\frac{1}{2\pi \sigma_n^2}\right)^M\exp\left\{-\frac{\Vert \boldsymbol{y}\Vert^2 + \Vert h\boldsymbol{s}\Vert^2}{2\sigma_n^2}\right\}.\]

The coherent detection problem in (\ref{eq7}) is simplified as:
\begin{equation}\label{eq8}
\begin{split}
\hat{k}_{\mathrm{e},1}, \hat{k}_{\mathrm{e},2}, \hat{k}_{\mathrm{o},1}, \hat{k}_{\mathrm{o},2} &=\mathrm{arg}\max_{k_\mathrm{e},k_\mathrm{o}} ~\mathrm{prob}\left(\boldsymbol{y}\vert \boldsymbol{s},h\right)\\
&=\mathrm{arg}\max_{k_\mathrm{e},k_\mathrm{o}} ~\Re\left\{\langle \boldsymbol{y},h\boldsymbol{s}\rangle\right\}.
\end{split}
\end{equation}

Considering that \(\overline{(\cdot)}\) evaluates the complex conjugate, \(\langle \boldsymbol{y},h\boldsymbol{s}\rangle\) can be simplified as:
\begin{equation}\label{eq9}
\begin{split}
\langle \boldsymbol{y},h\boldsymbol{s}\rangle & = \overline{h} \sum_{n=0}^{M-1}y(n)\overline{s}(n)\\
&= \overline{h} \sum_{n=0}^{M-1}y(n)\bigl(\overline{g}_1(n)c_\mathrm{d}(n)+ \overline{g}_2(n)c_\mathrm{u}(n)\bigr)\\
&= \overline{h} \left(\sum_{n=0}^{M-1}r_1(n)\overline{g}_1(n)+ \sum_{n=0}^{M-1}r_2(n)\overline{g}_2(n)\right)\\
&=  \overline{h} \left(\sum_{n=0}^{M-1}r_1(n)\overline{f}_{\mathrm{e},1}(n) + \sum_{n=0}^{M-1}r_1(n)\overline{f}_{\mathrm{o},1}(n)\right.\\
&~~~+ \left.\sum_{n=0}^{M-1}r_2(n)\overline{f}_{\mathrm{e},2}(n) + \sum_{n=0}^{M-1}r_2(n)\overline{f}_{\mathrm{o},2}(n)\right)\\
&= \overline{h} \bigl( \mathcal{R}_1(k_{\mathrm{e},1})+ \mathcal{R}_1(k_{\mathrm{o},1})+ \mathcal{R}_2(k_{\mathrm{e},2})+ \mathcal{R}_2(k_{\mathrm{o},2})\bigr),
\end{split}
\end{equation}
where \(r_1(n) = y(n)c_\mathrm{d}(n)\), and \(r_2(n)= y(n)c_\mathrm{u}(n)\). \(\mathcal{R}_1(k)\) and \(\mathcal{R}_2(k)\) is the DFT of \(r_1(n)\) and \(r_2(n)\), respectively. Moreover, \(R_1(k_{\mathrm{e}})\) and \( \mathcal{R}_1(k_{\mathrm{o}})\) is the DFT of \(r_1(n)\) evaluated at even and odd indexes, respectively, whereas \(\mathcal{R}_2(k_{\mathrm{e}})\) and \( \mathcal{R}_2(k_{\mathrm{o}})\) is the DFT of \(r_2(n)\) evaluated at even and odd indexes, respectively. Taking into account the simplification of \(\langle \boldsymbol{y},h\boldsymbol{s}\rangle\) in (\ref{eq8}), the detection problem in (\ref{eq8}) becomes:
\begin{equation}\label{eq10}
\begin{split}
\hat{k}_{\mathrm{e},1}, \hat{k}_{\mathrm{e},2}, \hat{k}_{\mathrm{o},1}, \hat{k}_{\mathrm{o},2} &=\mathrm{arg}\max_{k_\mathrm{e},k_\mathrm{o}} ~\Re\left\{\overline{h} \bigl(\mathcal{R}_1(k_{\mathrm{e},1})+ \mathcal{R}_1(k_{\mathrm{o},1})\right.\bigr.\\
&\left.\bigl.~~~~~~~~~~~~~~~~+ \mathcal{R}_2(k_{\mathrm{e},2})+ \mathcal{R}_2(k_{\mathrm{o},2})\bigr.)\right\}.
\end{split}
\end{equation}

The FSs evaluated in (\ref{eq10}) can also be dis-jointly estimated as:
\begin{equation}\label{eq11}
\begin{split}
\hat{k}_{\mathrm{e},1} &=\mathrm{arg}\max_{k_\mathrm{e}}~\Re\left\{\overline{h} \mathcal{R}_1(k_{\mathrm{e},1})\right\},\\
\hat{k}_{\mathrm{e},2} &=\mathrm{arg}\max_{k_\mathrm{e}}~\Re\left\{\overline{h} \mathcal{R}_2(k_{\mathrm{e},2})\right\},\\
\hat{k}_{\mathrm{o},1} &=\mathrm{arg}\max_{k_\mathrm{o}} ~\Re\left\{\overline{h} \mathcal{R}_1(k_{\mathrm{o},1})\right\},\\
\hat{k}_{\mathrm{o},2} &=\mathrm{arg}\max_{k_\mathrm{o}} ~\Re\left\{\overline{h} \mathcal{R}_1(k_{\mathrm{0},2})\right\}.\\
\end{split}
\end{equation}
%
%
\subsubsection{Non-Coherent Detection}
\begin{figure}[tb]\centering
\includegraphics[trim={0 0 0 0},clip,scale=1]{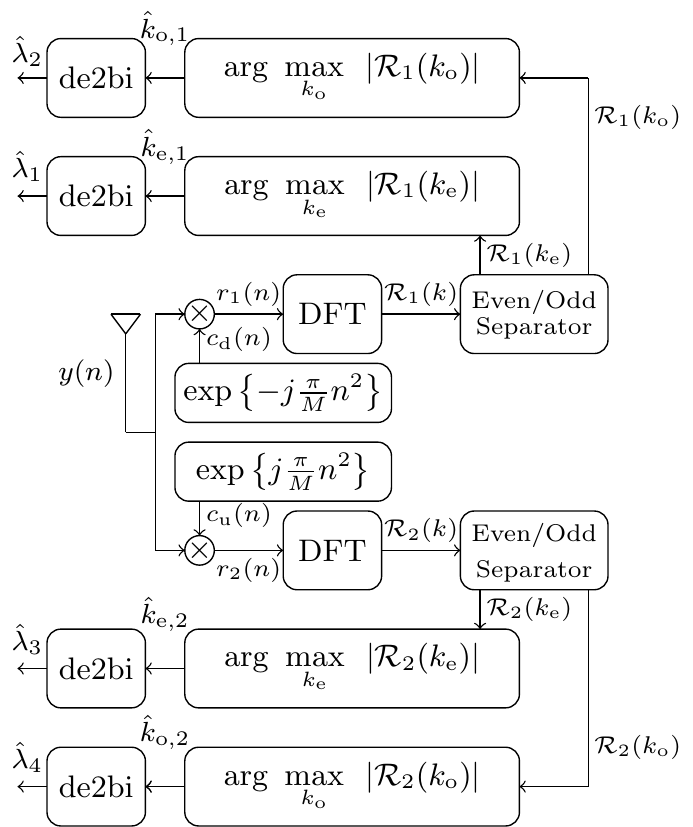}
  \caption{DM-TDM-CSS non-coherent detector architecture. }
\label{fig3}
\end{figure}
When the CSI is unavailable, the non-coherent detection mechanism can be used. It is more practical because its computational complexity is considerably lower than the coherent detection, which is better for low-power consumption and low-cost components in LPWANs. The non-coherent detector for DM-TDM-CSS is presented in Fig. \ref{fig3}. For non-coherent detection of DM-TDM-CSS, the FSs from the received symbols in a dis-joint fashion can be identified as: 
\begin{equation}\label{eq13}
\begin{split}
\hat{k}_{\mathrm{e},1} &=\mathrm{arg}\max_{k_\mathrm{e}}~\left\vert \mathcal{R}_1(k_{\mathrm{e},1})\right\vert,\\
\hat{k}_{\mathrm{e},2} &=\mathrm{arg}\max_{k_\mathrm{e}}~\left\vert \mathcal{R}_2(k_{\mathrm{e},2})\right\vert,\\
\hat{k}_{\mathrm{o},1} &=\mathrm{arg}\max_{k_\mathrm{o}} ~\left\vert \mathcal{R}_1(k_{\mathrm{o},1})\right\vert,\\
\hat{k}_{\mathrm{o},2} &=\mathrm{arg}\max_{k_\mathrm{o}} ~\left\vert \mathcal{R}_2(k_{\mathrm{0},2})\right\vert.\\
\end{split}
\end{equation}

From (\ref{eq13}), it is observed that the DFT of \(r_1(n)\) and \(r_2(n)\) is first evaluated, which yields \(\mathcal{R}_1(k)\) and \(R_2(k)\). Subsequently, the FS, \(\hat{k}_{\mathrm{e},1}\) and \(\hat{k}_{\mathrm{o},1}\) are determined using \(\mathcal{R}_1(k)\) by separating the even and odd frequency tones, respectively, whereas \(\hat{k}_{\mathrm{e},2}\) and \(\hat{k}_{\mathrm{o},2}\) are evaluated using \(\mathcal{R}_2(k)\) by isolating the even and odd frequency tones. 
\subsection{Orthogonality Analysis}
To analyze whether the DM-TDM-CSS symbols are orthogonal, we evaluate the inner product of the two distinct DM-TDM-CSS symbols, i.e., \(\boldsymbol{s}\) and \(\tilde{\boldsymbol{s}}= \left[\tilde{s}(0),\tilde{s}(1),\cdots, \tilde{s}(M-1)\right]^\mathrm{T}\) as \(\langle\boldsymbol{s},\tilde{\boldsymbol{s}}\rangle= \sum_{n=0}^{M-1} s(n) \overline{\tilde{s}}(n)\). The activated even, and odd FSs in the two chirped symbols of \(\boldsymbol{s}\) are \(k_{\mathrm{e},1}\), \(k_{\mathrm{e},2}\), \(k_{\mathrm{o},1}\), and \(k_{\mathrm{o},2}\). On the other hand, in \(\tilde{\boldsymbol{s}}\), the even and odd FSs in the two chirped symbols are \(\tilde{k}_{\mathrm{e},1}\), \(\tilde{k}_{\mathrm{e},2}\), \(\tilde{k}_{\mathrm{o},1}\), and \(\tilde{k}_{\mathrm{o},2}\). The following conditions must hold to determine if the DM-TDM-CSS symbols are orthogonal or not: (i) the even FS in the up-chirp and the down-chirp symbols, \(\hat{k}_{\mathrm{e},1}\),  and \(k_{\mathrm{e},2}\) in \(\boldsymbol{s}\) are different from the respective FS in \(\tilde{\boldsymbol{s}}\), \(\tilde{k}_{\mathrm{e},1}\), \(\tilde{k}_{\mathrm{e},2}\), i.e., \(k_{\mathrm{e},1} \neq \tilde{k}_{\mathrm{e},1}\), and \(k_{\mathrm{e},2} \neq \tilde{k}_{\mathrm{e},2}\); (ii) the same condition also holds for the odd FS of \(\boldsymbol{s}\) and \(\tilde{\boldsymbol{s}}\), i.e., \(k_{\mathrm{o},1} \neq \tilde{k}_{\mathrm{o},1}\), and \(k_{\mathrm{o},2} \neq \tilde{k}_{\mathrm{o},2}\). After some straightforward manipulation, the inner product \(\langle\boldsymbol{s},\tilde{\boldsymbol{s}}\rangle\) yields:
\begin{equation}\label{eq13}
\begin{split}
\langle\boldsymbol{s},\tilde{\boldsymbol{s}}\rangle &=\underbrace{\sum_{n=0}^{M-1}\exp\left\{j\frac{\pi}{M}\left(2k_1n+2n^2\right)\right\}}_{:=\tau_1} \\
&~~+\underbrace{\sum_{n=0}^{M-1}\exp\left\{j\frac{\pi}{M}\left(2k_2n+2n^2\right)\right\}}_{:=\tau_2} \\
&~~+\underbrace{\sum_{n=0}^{M-1}\exp\left\{j\frac{\pi}{M}\left(2k_3n-2n^2\right)\right\}}_{:=\tau_3} \\
&~~+\underbrace{\sum_{n=0}^{M-1}\exp\left\{j\frac{\pi}{M}\left(2k_4n-2n^2\right)\right\}}_{:=\tau_4},
\end{split}
\end{equation}
where \(k_1 = k_{\mathrm{e},1}-\tilde{k}_{\mathrm{e},2}\), \(k_2 = k_{\mathrm{o},1}-\tilde{k}_{\mathrm{o},2}\), \(k_3 = k_{\mathrm{e},2}-\tilde{k}_{\mathrm{e},1}\), and \(k_4 = k_{\mathrm{o},2}-\tilde{k}_{\mathrm{o},1}\). The closed-form expressions for \(\tau_1\), \(\tau_2\), \(\tau_3\), and \(\tau_4\) can be obtained using (\ref{eq2_apdx}) as in Appendix \ref{apdx}. Firstly, consider \(\tau_1\), for which \(a = 2\), \(b=2k_1\), and \(c = M\). To this end, we attain:
\begin{equation}\label{eq14}
\begin{split}
\tau_1&=\sqrt{\left\vert \frac{M}{2}\right\vert }\exp\left\{j\frac{\pi}{8M}\left(\vert 2M\vert -(2k_1)^2\right)\right\}\beta_1\\
\end{split}
\end{equation}
where
\begin{equation}\label{eq14a}
\begin{split}
\beta_1&=\sum_{n=0}^{1}\exp\left\{-j\frac{\pi}{2}\left(2k_1n +Mn^2\right)\right\}\\
&=1+\exp\left\{-j\frac{\pi}{2}\left( M +2k_1\right)\right\}\\
\end{split}
\end{equation}

Since \(M\) and \(\vert k_1 \vert\) are always even; therefore, \(\beta_1=2\) that leads to:
%
\begin{equation}\label{eq15}
\begin{split}
\tau_1=\alpha\exp\left\{-j\frac{\pi}{2M}k_1^2\right\} = \alpha\exp\left\{-j\frac{\pi}{2M}\left(k_{\mathrm{e},1}-\tilde{k}_{\mathrm{e},2}\right)^2\right\},
\end{split}
\end{equation}
where \[\alpha =2\sqrt{\frac{M}{2}}\exp\left\{j\frac{\pi}{4}\right\}.\]

Following the same steps as in (\ref{eq13}) and  (\ref{eq14}), \(\tau_2\), \(\tau_3\), and \(\tau_4\) are obtained as:
%
\begin{equation}\label{eq16}
\tau_2= \alpha \exp\left\{-j\frac{\pi}{2M}k_2^2\right\}= \alpha\exp\left\{-j\frac{\pi}{2M}\left(k_{\mathrm{o},1}-\tilde{k}_{\mathrm{o},2}\right)^2\right\},
\end{equation}
\begin{equation}\label{eq17}
\tau_3= \overline{\alpha}\exp\left\{j\frac{\pi}{2M}k_3^2\right\}= \overline{\alpha}\exp\left\{j\frac{\pi}{2M}\left(k_{\mathrm{e},2}-\tilde{k}_{\mathrm{e},1}\right)^2\right\},
\end{equation}
and 
\begin{equation}\label{eq18}
\tau_4= \overline{\alpha}\exp\left\{j\frac{\pi}{2M}k_4^2\right\}= \overline{\alpha}\exp\left\{j\frac{\pi}{2M}\left(k_{\mathrm{o},2}-\tilde{k}_{\mathrm{o},1}\right)^2\right\},
\end{equation}
respectively. Finally, the closed-form of (\ref{eq13}), i.e., \(\langle\boldsymbol{s},\tilde{\boldsymbol{s}}\rangle\) is given as:
\begin{equation}\label{eq19}
\begin{split}
\langle\boldsymbol{s},\tilde{\boldsymbol{s}}\rangle =\alpha\left(\theta_1+ \theta_2\right) + \overline{\alpha}\left(\theta_3+ \theta_4\right)
\end{split}
\end{equation}
where 
\begin{equation}\label{eq19a}
\theta_1 = \exp\left\{-j\frac{\pi}{2M}\left(k_{\mathrm{e},1}-\tilde{k}_{\mathrm{e},2}\right)^2\right\},
\end{equation}
\begin{equation}\label{eq19b}
\theta_2 = \exp\left\{-j\frac{\pi}{2M}\left(k_{\mathrm{o},1}-\tilde{k}_{\mathrm{o},2}\right)^2\right\},
\end{equation}
\begin{equation}\label{eq19c}
\theta_3 = \exp\left\{j\frac{\pi}{2M}\left(k_{\mathrm{e},2}-\tilde{k}_{\mathrm{e},1}\right)^2\right\},
\end{equation}
and
\begin{equation}\label{eq19d}
\theta_4 = \exp\left\{j\frac{\pi}{2M}\left(k_{\mathrm{o},2}-\tilde{k}_{\mathrm{o},1}\right)^2\right\},
\end{equation}
respectively. 

(\ref{eq19}) implies that simultaneously activating even and the odd FSs in the two multiplexed chirped symbols cause a loss of orthogonality between the two DM-TDM-CSS symbols. Precisely, the even FS of one chirped symbol induces interference for the even FS of the other chirped symbol and vice versa. The same is the case for the odd FSs, where the odd FS of one chirped symbol causes interference for the other chirped symbol. In the following section, we shall evaluate this interference quantitatively. 
\subsection{Interference Analysis}
The interference in the DM-TDM-CSS symbols can be determined by analyzing \(r_1(n)\) and \(r_2(n)\). In order to do so, we consider that \(y(n) = s(n) + w(n)\). In the following analysis, we shall treat \(r_1(n)\) and \(r_2(n)\) to determine the interference. Firstly, let us consider \(r_1(n)\), which is given as:
\begin{equation}\label{eq20}
\begin{split}
r_1(n) &= y(n)c_\mathrm{d}(n) \\
&= s(n)c_\mathrm{d}(n)+ \tilde{w}(n)\\
&= \bigl(g_1(n)c_\mathrm{u}(n)+g_2(n)c_\mathrm{d}(n)\bigr)c_\mathrm{d}(n)+ \tilde{w}(n)\\
&= g_1(n) + g_2(n)c_\mathrm{d}^2(n)+ \tilde{w}(n),
\end{split}
\end{equation}
where \(\tilde{w}(n) = w(n)c_\mathrm{d}(n)\) and \(c_\mathrm{u}(n)c_\mathrm{d}(n)= 1\). \(r_1(n)\) in (\ref{eq20}) can be re-written as:
\begin{equation}\label{eq21}
\begin{split}
r_1(n) &= \exp\left\{j\frac{2\pi}{M}k_{\mathrm{e},1}n\right\}+ \exp\left\{j\frac{2\pi}{M}k_{\mathrm{o},1}n\right\}\\
&~~+ \exp\left\{-j\frac{2\pi}{M}n^2\right\}\exp\left\{j\frac{2\pi}{M}k_{\mathrm{e},2}n\right\}\\
&~~+\exp\left\{-j\frac{2\pi}{M}n^2\right\}\exp\left\{j\frac{2\pi}{M}k_{\mathrm{o},2}n\right\}+ \tilde{w}(n).
\end{split}
\end{equation}

Taking \(M\)-order DFT of \(r_1(n)\) yields \(\mathcal{R}_1(k)\), i.e.,
\begin{equation}\label{eq22}
\begin{split}
\mathcal{R}_1(k) &=\underbrace{\sum_{n=0}^{M-1} \exp\left\{j\frac{2\pi}{M}\tilde{k}_1n\right\}}_{:=\kappa_1}+\underbrace{\sum_{n=0}^{M-1} \exp\left\{j\frac{2\pi}{M}\tilde{k}_2n\right\}}_{:=\kappa_2} \\
&~~+ \underbrace{\sum_{n=0}^{M-1}\exp\left\{j\frac{\pi}{M}\left(2\tilde{k}_3n-2n^2\right)\right\}}_{:=\kappa_3}\\
&~~+ \underbrace{\sum_{n=0}^{M-1}\exp\left\{j\frac{\pi}{M}\left(2\tilde{k}_4n-2n^2\right)\right\}}_{:=\kappa_4}+\tilde{W}(k),
\end{split}
\end{equation}
where \(\tilde{k}_1 = k_{\mathrm{e},1}-k\), \(\tilde{k}_2 = k_{\mathrm{o},1}-k\), \(\tilde{k}_3 = k_{\mathrm{e},2}-k\), and \(\tilde{k}_4 = k_{\mathrm{o},2}-k\). 

Now considering \(k \in \tilde{k}_\mathrm{e}\) for (\ref{eq22}), we ascertain that \(\kappa_1 = M\) when \(\tilde{k}_1 = k_{\mathrm{e},1}\), \(\kappa_2 = 0\) because \(\tilde{k}_2 \neq 0\). Moreover, using the generalized quadratic Gauss sum as given in (\ref{eq2_apdx}) of Appendix \ref{apdx}, \(\kappa_3\) is given as:
\begin{equation}
\kappa_3 = \frac{\overline{\alpha}}{2}\exp\left\{j\frac{\pi}{2M}\tilde{k}_3^2\right\}\left(1+\exp\left\{j\frac{\pi}{2}\left(M+2\tilde{k}_3\right)\right\}\right).
\end{equation} 

If \(M\) is a power of \(2\) (which in general it is) and if \(k \in \tilde{k}_\mathrm{e}\), then \(\exp\left\{j\frac{\pi}{2}\left(M+2\tilde{k}_3\right)\right\}=1\), which leads to 
\begin{equation}
\kappa_3 = \overline{\alpha}\exp\left\{j\frac{\pi}{2M}\tilde{k}_3^2\right\}.
\end{equation} 

Solving \(\kappa_4\) using (\ref{eq2_apdx}) yields: 
\begin{equation}
\kappa_4 = \frac{\overline{\alpha}}{2}\exp\left\{j\frac{\pi}{2M}\tilde{k}_4^2\right\}\left(1+\exp\left\{j\frac{\pi}{2}\left(M+2\tilde{k}_4\right)\right\}\right).
\end{equation} 

Since \(\tilde{k}_4\) is always odd; therefore, \(\exp\left\{j\frac{\pi}{2}\left(M+2\tilde{k}_4\right)\right\}=-1\) resulting in \(\kappa_4= 0\). It is important to note that \(\kappa_2 = \kappa_4 = 0\) implies that the odd FS of one chirped symbol does not cause any interference with the even FS of the other. Thus, for \(k \in \tilde{k}_\mathrm{e}\), the output of the DFT when \(\tilde{k}_\mathrm{e} = k_{\mathrm{e},1}\) results in
\begin{equation}\label{eq23}
\begin{split}
\mathcal{R}_1(k_{\mathrm{e},1}) = \underbrace{M}_{\mathrm{signal}} + \underbrace{\overline{\alpha}\exp\left\{j\frac{\pi}{2M}(k_{\mathrm{e},2}-k_{\mathrm{e},1})^2\right\}}_{\mathrm{Interference}}+\tilde{W}(k_{\mathrm{e},1}).
\end{split}
\end{equation}

Performing similar analysis as done for (\ref{eq23}) considering \(k \in \tilde{k}_\mathrm{o}\) and \(\tilde{k}_\mathrm{o} = k_{\mathrm{o},1}\) yields
\begin{equation}\label{eq24}
\mathcal{R}_1(k_{\mathrm{o},1}) = \underbrace{M}_{\mathrm{signal}} + \underbrace{\overline{\alpha}\exp\left\{j\frac{\pi}{2M}(k_{\mathrm{o},2}-k_{\mathrm{o},1})^2\right\}}_{\mathrm{Interference}}+\tilde{W}(k_{\mathrm{o},1}).
\end{equation}

Note that while evaluating (\ref{eq24}), it is observed that \(\kappa_1 = \kappa_3=0\), implying that even FSs of one chirped symbol does not interfere with the odd FS of the other chirped symbol.

%
%

\(r_2(n)\) is attained as
\begin{equation}\label{eq25}
\begin{split}
r_2(n) &= g_1(n)c_\mathrm{u}^2(n) + g_2(n)+ \hat{\bar{w}}(n)\\
&= \exp\left\{j\frac{2\pi}{M}n^2\right\}\exp\left\{j\frac{2\pi}{M}k_{\mathrm{e},1}n\right\}\\
&+\exp\left\{j\frac{2\pi}{M}n^2\right\}\exp\left\{j\frac{2\pi}{M}k_{\mathrm{o},1}n\right\}\\
&+\exp\left\{j\frac{2\pi}{M}k_{\mathrm{e},2}n\right\}+\exp\left\{j\frac{2\pi}{M}k_{\mathrm{o},2}n\right\}+ \hat{\bar{w}}(n),\\
\end{split}
\end{equation}
where \(\hat{\bar{w}}(n)(n) = w(n)c_\mathrm{u}(n)\). \(M\)-order DFT of \(r_2(n)\) results in:
\begin{equation}\label{eq26}
\begin{split}
\mathcal{R}_2(k) &=\underbrace{\sum_{n=0}^{M-1} \exp\left\{j\frac{\pi}{M}\left(2\tilde{k}_1n + 2n^2\right)\right\}}_{:=\kappa_5}\\
&~~+\underbrace{\sum_{n=0}^{M-1} \exp\left\{j\frac{\pi}{M}\left(2\tilde{k}_2n + 2n^2\right)\right\}}_{:=\kappa_6} \\
&~~+ \underbrace{\sum_{n=0}^{M-1} \exp\left\{j\frac{\pi}{M}2\tilde{k}_3n\right\}}_{:=\kappa_7}+ \underbrace{\sum_{n=0}^{M-1} \exp\left\{j\frac{\pi}{M}2\tilde{k}_4n\right\}}_{:=\kappa_8}\\
&~~+\hat{\bar{W}}(k).
\end{split}
\end{equation}

For \(k \in \tilde{k}_\mathrm{e}\) when \(\tilde{k}_\mathrm{e} = k_{\mathrm{e},2}\), then using (\ref{eq2_apdx}), we attain \(\kappa_5 = \alpha\exp\left\{-j\frac{\pi}{2M}\left(k_{\mathrm{e},1}-\tilde{k}_\mathrm{e}\right)^2\right\}\), \(\kappa_6 = 0\), \(\kappa_7 = M\), and \(\kappa_8 = 0\) that leads to 
\begin{equation}\label{eq27}
\mathcal{R}_2(k_{\mathrm{e},2}) = \underbrace{M}_{\mathrm{signal}} + \underbrace{\alpha\exp\left\{-j\frac{\pi}{2M}(k_{\mathrm{e},1}-k_{\mathrm{e},2})^2\right\}}_{\mathrm{Interference}}+\hat{\bar{W}}(k_{\mathrm{e},2}).
\end{equation}

Similarly, for \(k \in \tilde{k}_\mathrm{e}\), we have \(\kappa_5 = 0\), \(\kappa_6 = \alpha\exp\left\{-j\frac{\pi}{2M}\left(k_{\mathrm{o},1}-\tilde{k}_\mathrm{o}\right)^2\right\}\), \(\kappa_7 = 0\), and \(\kappa_8 = M\) resulting in
\begin{equation}\label{eq28}
\mathcal{R}_2(k_{\mathrm{o},2}) = \underbrace{M}_{\mathrm{signal}} + \underbrace{\alpha\exp\left\{-j\frac{\pi}{2M}(k_{\mathrm{o},1}-k_{\mathrm{o},2})^2\right\}}_{\mathrm{Interference}}+\hat{\bar{W}}(k_{\mathrm{o},2}).
\end{equation}
%


From (\ref{eq23}) and (\ref{eq24}), we can observe that the activated FSs of the second chirped symbol, i.e., \(k_{\mathrm{e},2}\) and \(k_{\mathrm{o},2}\), cause interference when the activated FSs of the first chirp symbol, i.e., \(k_{\mathrm{e},1}\) and \(k_{\mathrm{o},1}\) are to be determined. We can draw similar conclusions from (\ref{eq27}) and (\ref{eq28}) that the FSs of the first chirped symbol, \(k_{\mathrm{e},1}\) and \(k_{\mathrm{o},1}\), cause interference when we need to determine the FSs of the second chirped symbol, \(k_{\mathrm{e},2}\) and \(k_{\mathrm{o},2}\). 

In addition, we can also explicitly obtain the expression for SIR, \(~\gamma\) as we have both the signal power and the interference power. Since, the interference power for \(\mathcal{R}_1(\tilde{k}_\mathrm{e}) \), \(\mathcal{R}_1(\tilde{k}_\mathrm{o}) \), \(\mathcal{R}_2(\tilde{k}_\mathrm{e}) \), and \(\mathcal{R}_2(\tilde{k}_\mathrm{o}) \) is the same, \(\gamma\) is evaluated as:
\begin{equation}\label{eq29}
\gamma = \frac{M^2}{\left\vert \alpha_3\exp\left\{j\frac{\pi}{M}(k_{\mathrm{e},2}-\tilde{k}_\mathrm{e})^2\right\} \right\vert^2} = \frac{M^2}{2M} = \frac{M}{2}.
\end{equation}

Notice that for the evaluation of \(\gamma\), we have used the interference power of \(\mathcal{R}_1(\tilde{k}_\mathrm{e}) \); however, as aforementioned, the interference power for \(\mathcal{R}_1(\tilde{k}_\mathrm{o}) \), \(\mathcal{R}_2(\tilde{k}_\mathrm{e}) \), and \(\mathcal{R}_2(\tilde{k}_\mathrm{o}) \) is also the same. Consequently, we will always obtain the same result for \(\gamma\). Furthermore, from (\ref{eq29}), we gather that the interference vanishes away with increasing \(M\), i.e., for higher \(\lambda\). 

We can also define signal-to-interference plus noise (SINR), \(\Gamma\) as:
\begin{equation}\label{eq30}
\Gamma = \frac{M^2}{2M+ \sigma_n^2} = \frac{M}{2+\frac{\sigma_n^2}{M}}.
\end{equation}

Again we observe that \(\Gamma\) also increases with an increase in \(M\), i.e., for higher \(\lambda\).
\section{Numerical Results and Discussions}
In this section, we compare the performance of the proposed DM-TDM-CSS with other classical counterparts in the state-of-the-art. We consider all the approaches in the literature that can transmit nearly \(2\lambda\) or higher number of bits per symbol of duration \(T_\mathrm{s}\). The schemes include IQ-CSS, TDM-CSS, and IQ-TDM-CSS. In addition, LoRa has also been considered a benchmark for comparison. To compare the proposed DM-TDM-CSS with other schemes, we consider the SE versus EE performance, BER performance in an AWGN and fading channels, and BER performance considering phase and frequency offsets. It is highlighted that we consider the non-coherent detection for IQ-CSS as proposed in \cite{iqcim}.
\subsection{Spectral Efficiency of DM-TDM-CSS}
In DM-TDM-CSS, four different bit sequences of length \(\lambda-1\) are used per symbol; therefore, the total number of bits transmitted per symbol of duration \(T_\mathrm{s}\) is \(4\lambda-4\). Note that the number of bits transmitted per DM-TDM-CSS symbol is \(4\) bits less than that of IQ-TDM-CSS; however, the advantages of DM-TDM-CSS will become evident in the following sections. With the given number of bits transmitted per symbol, the data rate SE of DM-TDM-CSS is given by \(R = \sfrac{(4\lambda-4)}{T_\mathrm{s}}\) bits/s/Hz. Moreover, considering that \(B = \sfrac{M}{T_\mathrm{s}}\), the SE of DM-TDM-CSS is 
\begin{equation}
\eta = \frac{R}{B} = \frac{4\lambda-4}{M}.
\end{equation}

The spectral efficiencies of other schemes considered in this article are presented in Table \ref{tab1}.
\begingroup
\setlength{\tabcolsep}{6pt} 
\renewcommand{\arraystretch}{1.3} 
\begin{table}[tb]
\caption{Spectral efficiencies of different CSS schemes}
\centering
\begin{tabular}{c||c}
\hline
\hline
\textbf{CSS Scheme} & \textbf{SE  (bits/s/Hz)}\\
\hline
\hline
LoRa& \(\frac{\lambda}{M}\)\\
\hline
IQ-CSS&\(\frac{2\lambda}{M}\)\\
\hline
TDM-CSS&\(\frac{2\lambda}{M}\)\\
\hline
IQ-TDM-CSS&\(\frac{4\lambda}{M}\)\\
\hline
DM-CSS&\(\frac{2\lambda+1}{M}\) \\
\hline 
\hline
\end{tabular}
\label{tab1}
\end{table}
\endgroup
\subsection{Spectral Efficiency versus Energy Efficiency Performance}
In this section, we evaluate and compare the SE versus EE performance of the proposed DM-TDM-CSS with other state-of-the-art schemes. In order to evaluate the performance at a given SE, we attain the EE by evaluating \(E_\mathrm{b}/N_0=\frac{E_\mathrm{s}T_\mathrm{s}}{\eta N_0}\) required to reach a BER of \(10^{-3}\). On the other hand, the SE is changed by varying \(\lambda = \llbracket 6,12\rrbracket\). This performance metric is evaluated for all the considered schemes in the AWGN channel. Moreover, the performance is considered separately for coherent and non-coherent detection mechanisms. 
\begin{figure}[tb]\centering
\includegraphics[trim={20 0 0 0},clip,scale=0.7]{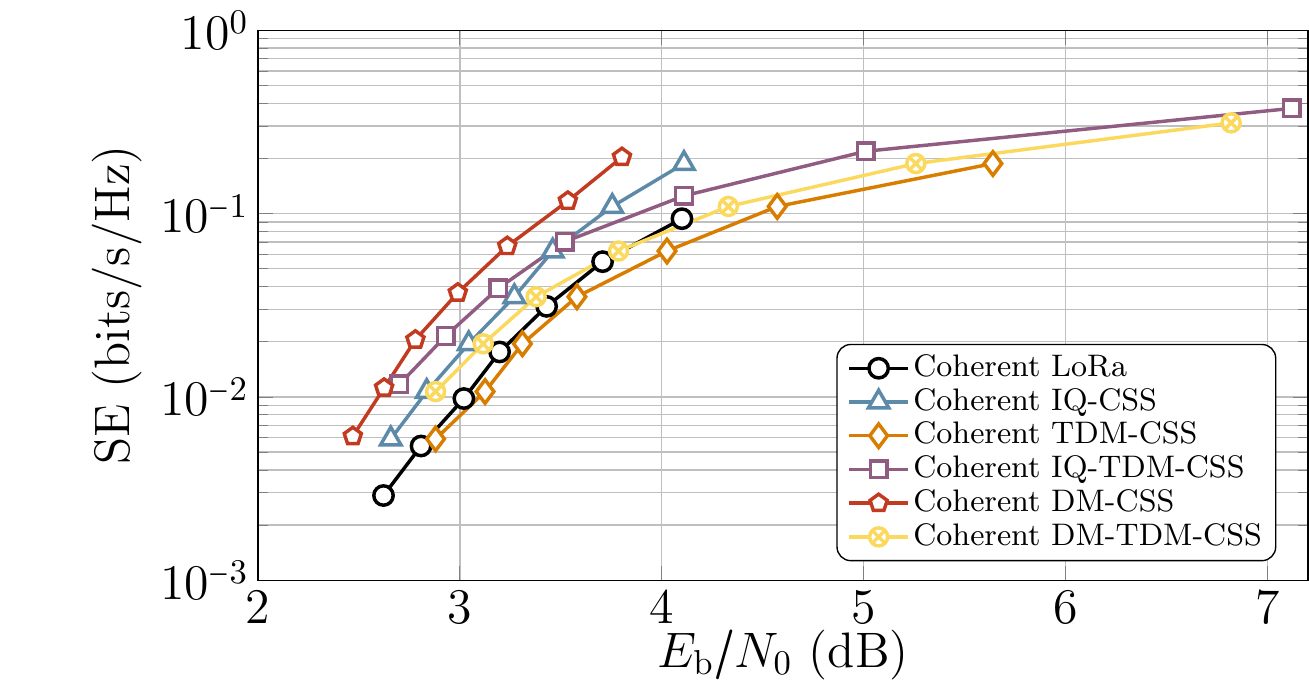}
  \caption{SE versus EE performance considering coherent detection and target BER of \(10^{-3}\) in the AWGN channel.}
\label{fig4}
\end{figure}
\begin{figure}[tb]\centering
\includegraphics[trim={20 0 0 0},clip,scale=0.7]{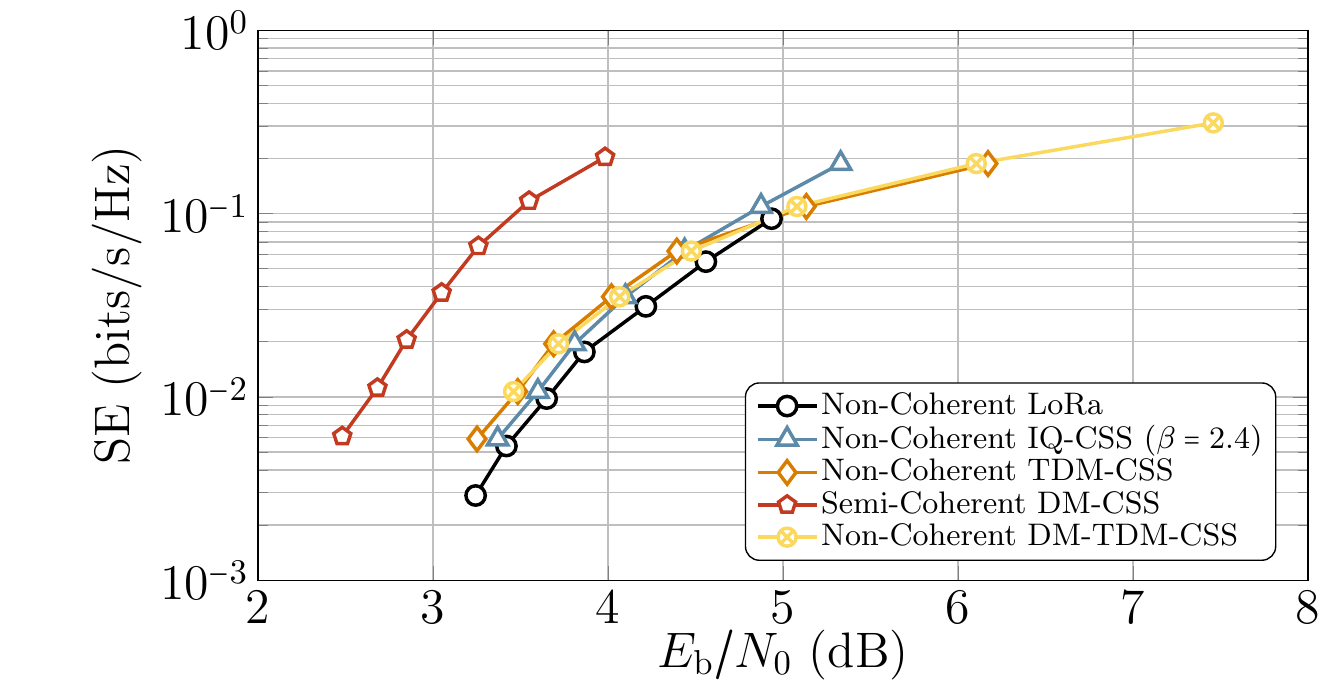}
  \caption{SE versus EE performance considering non-coherent detection and target BER of \(10^{-3}\) in the AWGN channel.}
\label{fig5}
\end{figure}
From Fig. \ref{fig4}, which illustrates SE versus EE performance considering coherent detection mechanism, we observe that the performance of DM-CSS is best among all the approaches. Another aspect that can be observed is that IQ-TDM-CSS can achieve the maximum achievable SE, whereas the maximum achievable SE of DM-TDM-CSS is marginally less than that of IQ-TDM-CSS. For \(\lambda>9\), the SE versus EE performance of IQ-TDM-CSS approaches that of DM-CSS, making it a desirable alternative to other approaches when coherent detection is employed. On the other hand, the performance of DM-TDM-CSS is close to that of TDM-CSS. IQ-CSS also performs better than the proposed DM-TDM-CSS. From Fig. \ref{fig4}, it seems that the proposed approach may not be an acceptable alternative to already proposed approaches in the literature. However, when we consider the performance of the same schemes considering non-coherent detection, as illustrated in Fig. \ref{fig5}, it is evident that DM-TDM-CSS is one of the best approaches regarding SE versus EE performance. The reasons are (i) DM-CSS only offers semi-coherent detection because the evaluation of PSs needs a coherent maximum likelihood detection mechanism at the receiver; therefore, the overall detection complexity in DM-CSS is high; and (ii) IQ-TDM-CSS symbols cannot be detected using non-coherent detection, which is a significant limitation. 
\subsection{BER Performance in AWGN Channel}
\begin{figure}[tb]\centering
\includegraphics[trim={26 0 0 0},clip,scale=0.87]{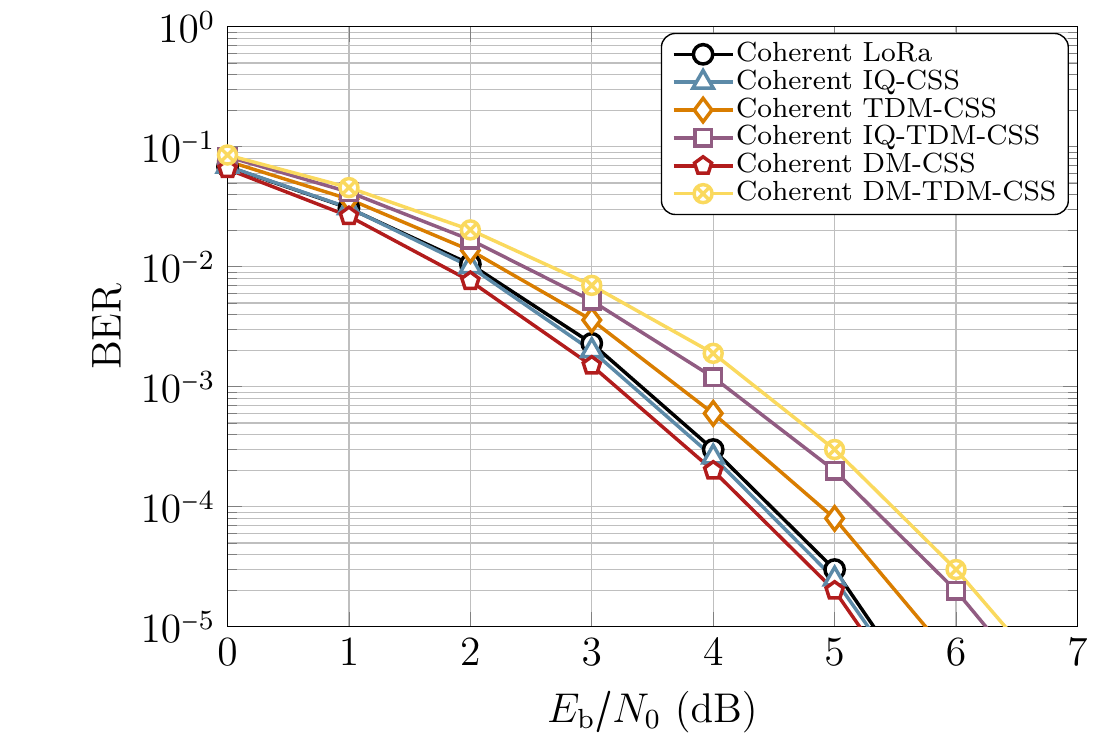}
  \caption{BER performance considering coherent detection in AWGN channel for \(\lambda = 8\).}
\label{fig6}
\end{figure}
\begin{figure}[tb]\centering
\includegraphics[trim={26 0 0 0},clip,scale=0.87]{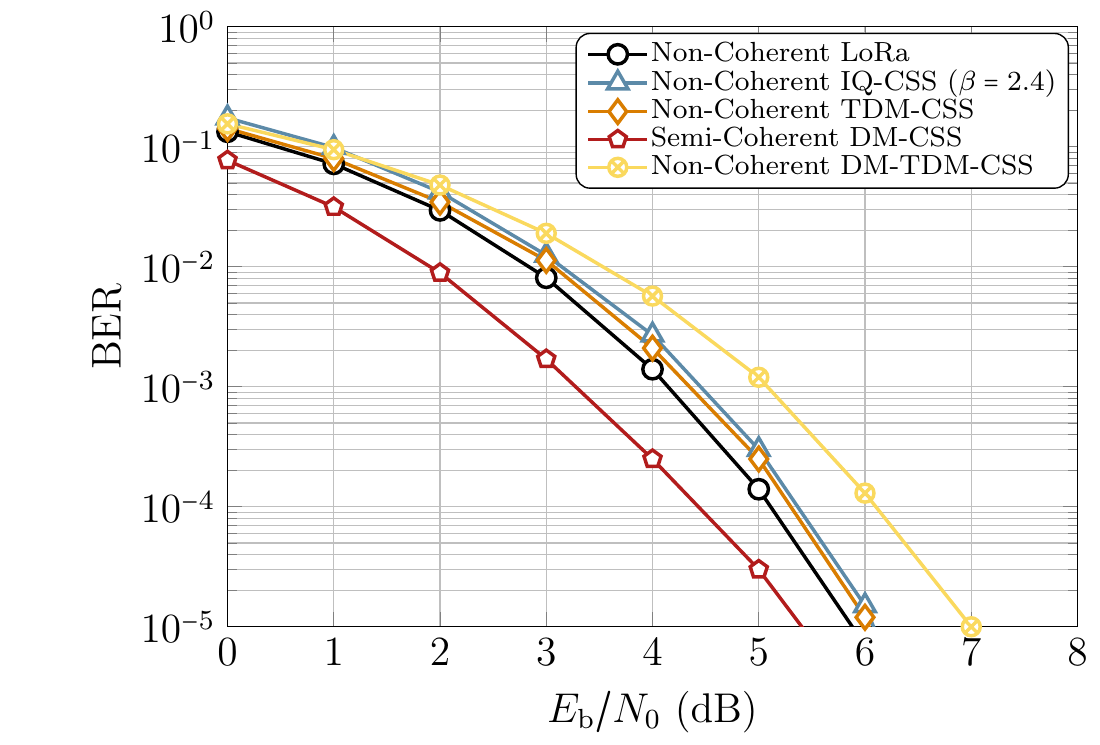}
  \caption{BER performance considering coherent detection in AWGN channel for \(\lambda = 8\).}
\label{fig7}
\end{figure}
Fig. \ref{fig6} and Fig. \ref{fig7} illustrate the BER performance of the proposed DM-TDM-CSS scheme and compare it with other alternatives considering coherent and non-coherent detection, respectively. The BER performance is obtained considering an AWGN for \(\lambda = 8\). For \(\lambda=8\), the spectral efficiencies of the considered schemes are LoRa: \(0.0312\) bits/s/Hz, IQ-CSS: \(0.0625\) bits/s/Hz, TDM-CSS: \(0.0625\) bits/s/Hz, IQ-TDM-CSS: \(0.125\) bits/s/Hz, DM-CSS: \(0.0664\) bits/s/Hz, and DM-TDM-CSS: \(0.1093\) bits/s/Hz. With the given spectral efficiencies, DM-TDM-CSS offers approximately \(250\%\) increase in SE over LoRa, \(75\%\) increase in SE over IQ-CSS and TDM-CSS, and \(65\%\) over DM-CSS. However, the SE of DM-TDM-CSS is approximately \(14\%\) less relative to IQ-TDM-CSS. 

The BER performance coherent detection (cf. Fig. \ref{fig6}) illustrates that the BER of DM-TDM-CSS is marginally higher relative to IQ-TDM-CSS, i.e., DM-TDM-CSS requires approximately \(0.2\) dB higher \(E_\mathrm{b}/N_0\) to attain a BER of \(10^{-3}\). As expected, both DM-TDM-CSS and IQ-TDM-CSS would have a higher BER because they offer higher spectral efficiencies relative to other schemes. It can also be observed that the performance of DM-CSS performs the best for coherent detection. 

The added value of DM-TDM-CSS becomes evident when we analyze the BER performance considering non-coherent detection (cf. Fig.  \ref{fig7}). IQ-TDM-CSS does not allow non-coherent detection, which makes it less advantageous. On the other hand, DM-TDM-CSS allows non-coherent detection. It can be observed that the DM-TDM-CSS requires approximately \(0.5\) dB higher \(E_\mathrm{b}/N_0\) to attain a BER of \(10^{-3}\) relative to IQ-CSS and TDM-CSS. However, considering that the SE increase is \(65\%\), the increase is insignificant, which makes DM-TDM-CSS a solid alternative to the classical approaches. 
\subsection{BER Performance in Fading Channel}
\begin{figure}[tb]\centering
\includegraphics[trim={26 0 0 0},clip,scale=0.87]{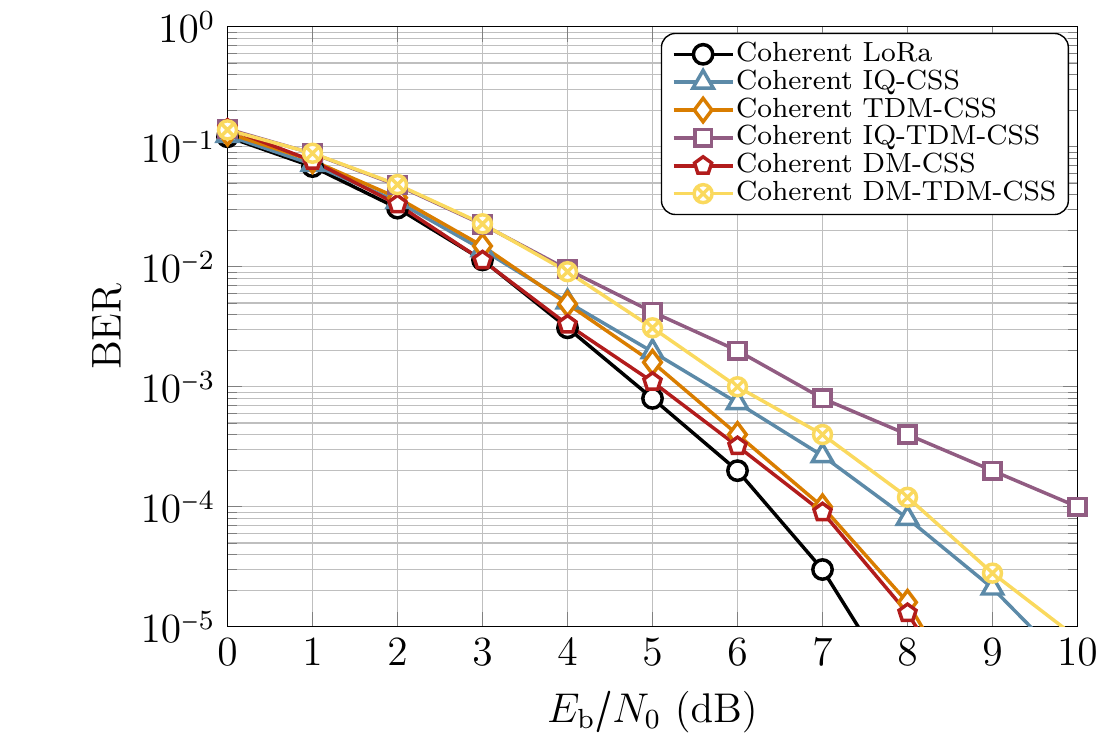}
  \caption{BER performance considering coherent detection in fading channel for \(\lambda = 8\).}
\label{fig8}
\end{figure}
\begin{figure}[tb]\centering
\includegraphics[trim={26 0 0 0},clip,scale=0.87]{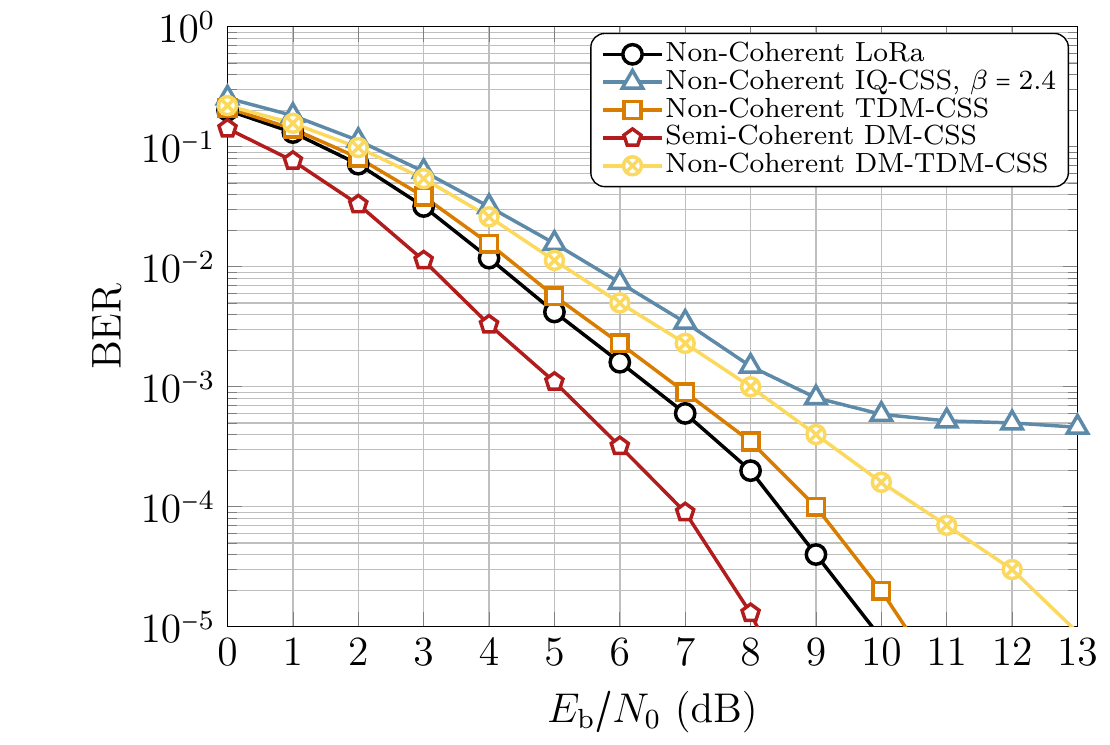}
  \caption{BER performance considering non-coherent detection in fading channel for \(\lambda = 8\).}
\label{fig9}
\end{figure}
In this section, we consider a frequency-selective \(2\)-tap fading channel having an impulse response of \(h(n) = \sqrt{1-\rho}\delta (nT) + \sqrt{\rho}\delta(nT-T)\), where \(T\) is the sampling duration and \(0\leq \rho  \leq 1\). The results are also depicted in Fig. \ref{fig2}, where \(\rho = 0.2\). The BER performance of the considered approaches considering coherent and non-coherent detection is illustrated in Fig. \ref{fig7} and Fig. \ref{fig8}. 

From the BER performance considering coherent detection (cf. Fig. \ref{fig8}), we can observe that, unlike the BER performance in the AWGN channel, the BER performance of DM-TDM-CSS is reasonably better than that of IQ-TDM-CSS. The reason is that DM-TDM-CSS does not transmit any information in the in-phase and quadrature components like IQ-TDM-CSS, which makes the latter approach more susceptible to frequency selective fading. On the other hand, the performance of DM-TDM-CSS is almost similar to that of IQ-CSS. It is essential to highlight that for a similar performance, the SE of DM-TDM-CSS is \(65\%\) higher than IQ-CSS.

The BER performance in fading channel considering non-coherent detection, as depicted in Fig. \ref{fig9}, illustrates that the performance of DM-TDM-CSS is better than IQ-CSS. It is recalled that IQ-TDM-CSS cannot be detected non-coherently. On the other hand, DM-TDM-CSS requires \(1\) dB higher \(E_\mathrm{b}/N_0\) to attain a BER of \(10^{-3}\) relative to TDM-CSS; however, the former approach's SE is \(65\%\) higher than that of the latter approach. Moreover, DM-CSS only offers a semi-coherent detection; therefore, its performance is better than the remaining counterparts. 
\subsection{BER Performance Considering Phase Offset}
\begin{figure}[tb]\centering
\includegraphics[trim={26 0 0 0},clip,scale=0.87]{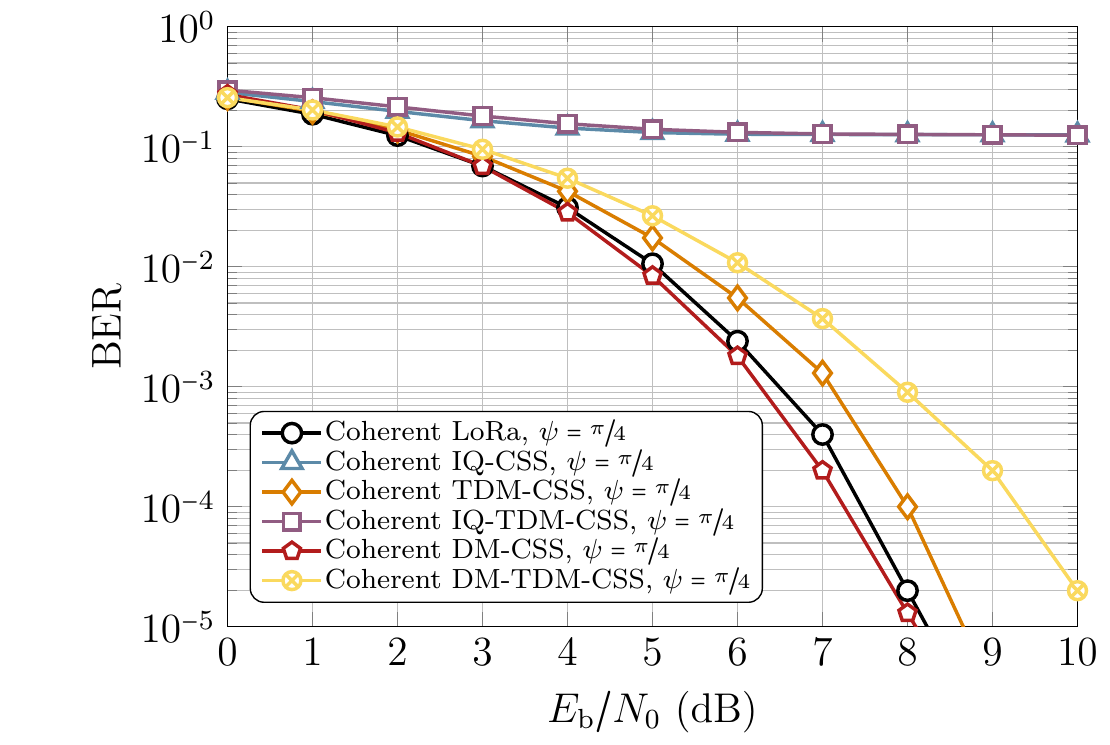}
  \caption{BER performance for different schemes considering coherent detection, \(\psi=\sfrac{\pi}{4}\) radians, AWGN channel, and \(\lambda = 8\).}
\label{fig10}
\end{figure}
\begin{figure}[tb]\centering
\includegraphics[trim={26 0 0 0},clip,scale=0.87]{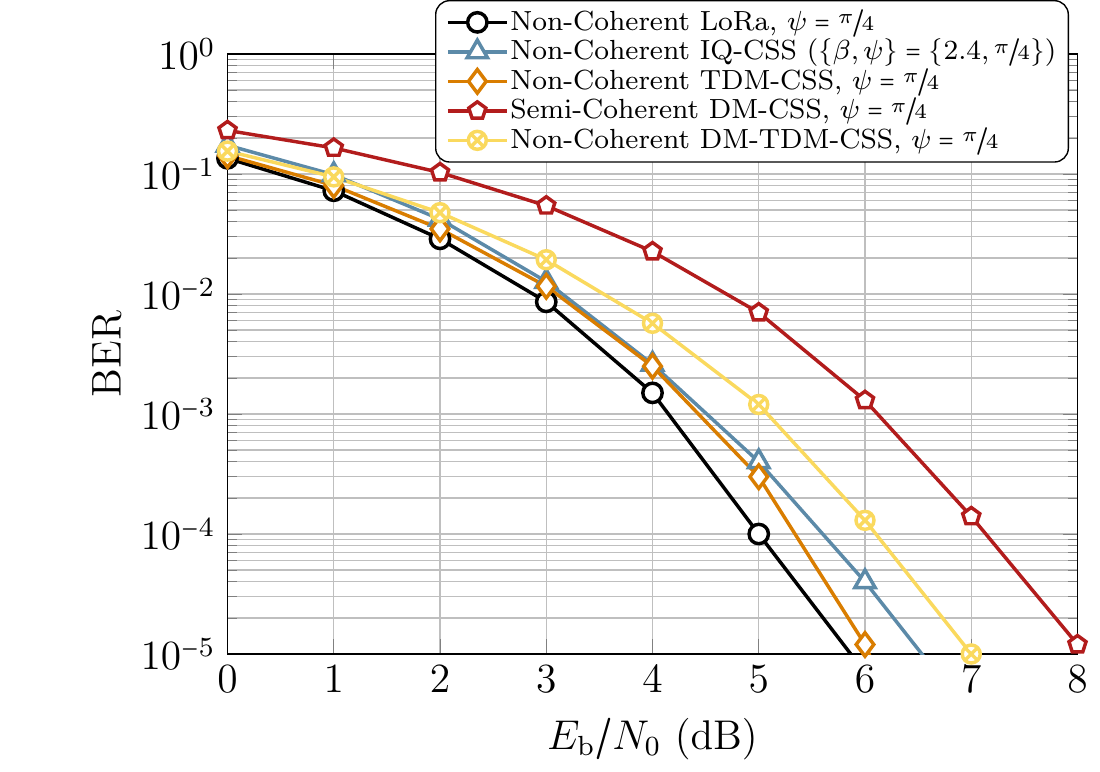}
  \caption{BER performance for different schemes considering non-coherent detection, \(\psi=\sfrac{\pi}{4}\) radians, AWGN channel, and \(\lambda = 8\).}
\label{fig11}
\end{figure}
In this section, we analyze the performance of all the schemes considering PO, which is expected to exist in low-cost devices. To this end, the received symbol corrupted by PO and AWGN is given as:
\begin{equation}\label{po_eq}
y(n) = \exp\{j\psi\} s(n) + w(n),
\end{equation}
where \(\psi\) is the PO. We evaluate the performance of the considered schemes for coherent and non-coherent detection and consider a PO of \(\psi = \sfrac{\pi}{4}\) radians and \(\lambda = 8\).

Fig. \ref{fig10} depicts the BER performance of the considered schemes using coherent detection and a PO of \(\psi = \sfrac{\pi}{4}\) radians. The performance of IQ-CSS and IQ-TDM-CSS is severely affected due to the PO because these schemes incorporate additional information in the IQ components. On the other hand, the performance of DM-TDM-CSS remains essentially robust against the PO. It can also be seen that the BER performance of IQ-CSS and IQ-TDM-CSS was better than the proposed DM-TDM-CSS in the AWGN channel (cf. Fig. \ref{fig6}); however, in the presence of distortions, the performance degrades severely. It is also accentuated that apart from DM-TDM-CSS, no other scheme can transmit \(4\lambda-4\) bits per symbol and is also robust against the PO.

Fig. \ref{fig11} portrays the BER performance of schemes that employ non-coherent detection and a PO of \(\psi = \sfrac{\pi}{4}\) radians. It may be noticed that the semi-coherently detected DM-CSS that was performing the best in the AWGN and fading channels is severely influenced by the PO. Moreover, the performance of non-coherently detected TDM-CSS and IQ-CSS remains better than DM-TDM-CSS; however, these schemes possess half of the SE of DM-TDM-CSS. Consequently, DM-TDM-CSS is the only approach that transmits \(4\lambda-4\) bits per symbol while also employing non-coherent detection. 
\subsection{BER Performance Considering Frequency Offset}
\begin{figure}[tb]\centering
\includegraphics[trim={26 0 0 0},clip,scale=0.87]{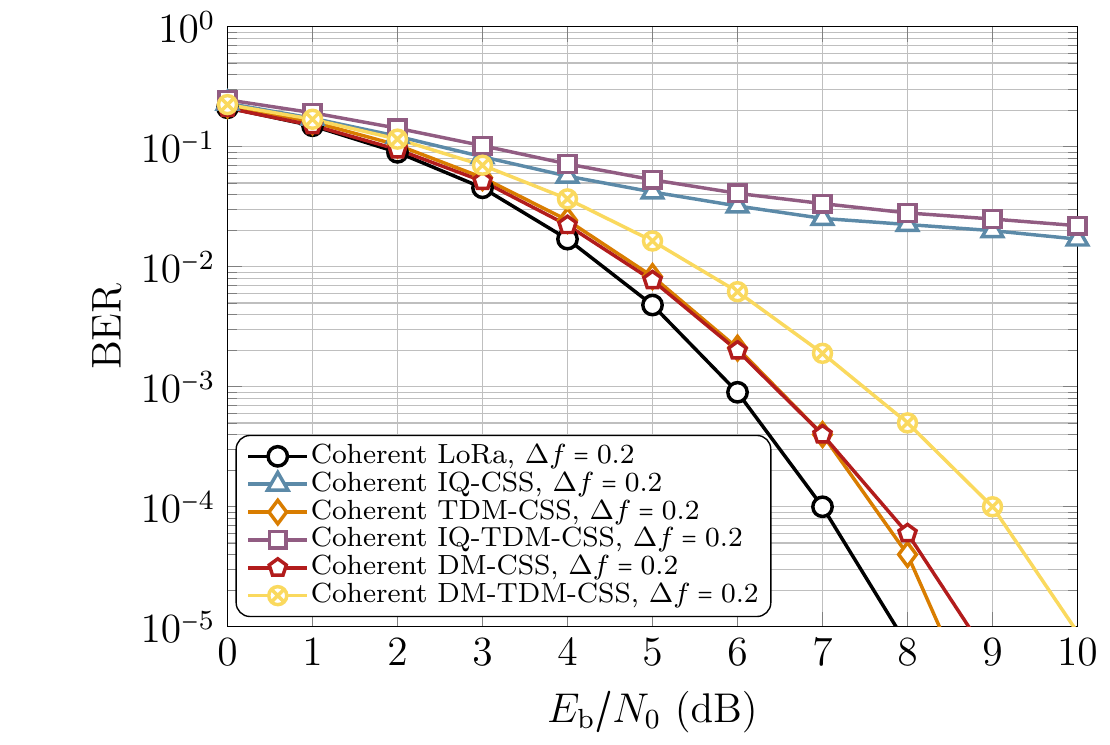}
  \caption{BER performance for different schemes considering coherent detection, \(\Delta f = 0.2\) Hz, AWGN channel, and \(\lambda = 8\).}
\label{fig12}
\end{figure}
\begin{figure}[tb]\centering
\includegraphics[trim={26 0 0 0},clip,scale=0.87]{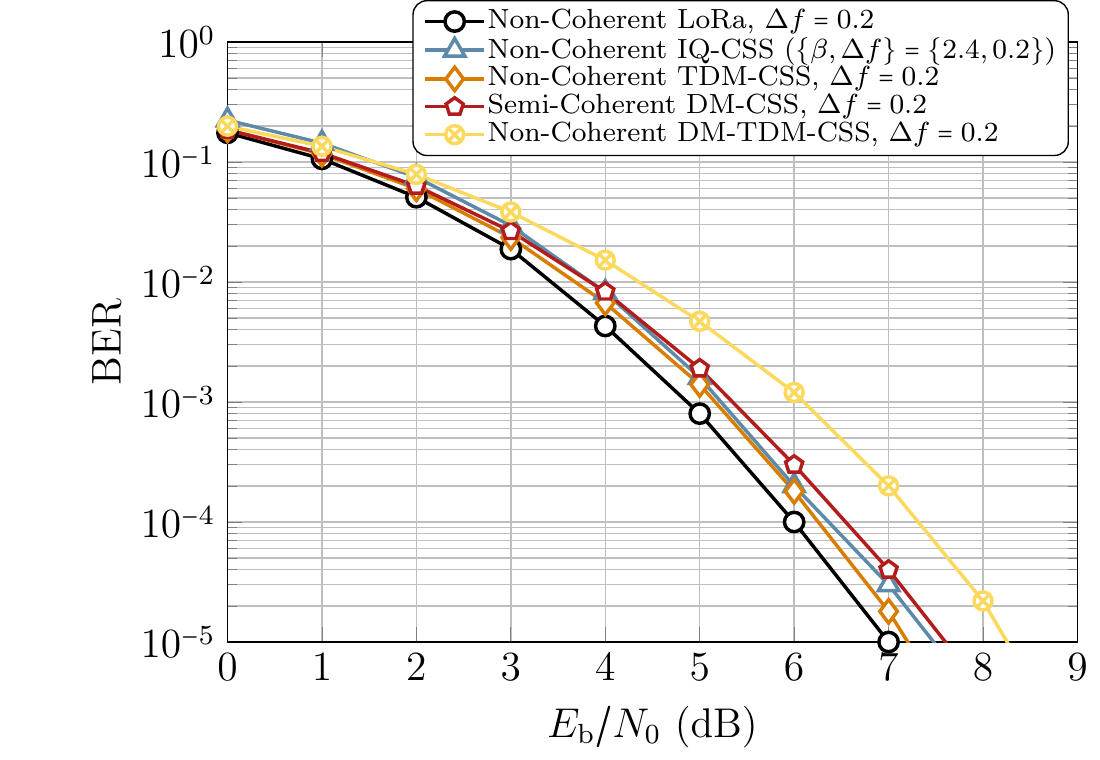}
  \caption{BER performance for different schemes considering non-coherent detection, \(\Delta f = 0.2\) Hz, AWGN channel, and \(\lambda = 8\).}
\label{fig13}
\end{figure}
In this section, we investigate the BER performance in the presence of  FO. The carrier FO linearly accumulates phase rotations from one symbol to another. In this case, the received symbol incorporating the impact of FO is 
\begin{equation}
y(n) = \exp\left\{\frac{j2\pi \Delta f n}{M}\right\} s(n) + w(n), 
\end{equation}
where \(\Delta f\) is the FO in Hz\footnote{\(\Delta f\) could also be seen as residual FO because IoT modems (for example, Bluetooth) normally implement a carrier frequency offset compensator before demodulation, thus, after compensation, \(\Delta f\) will reflect the residual FO.}. To evaluate the BER performance, we consider FO of  \(\Delta f = 0.2\) Hz, \(\lambda = 8\), and AWGN channel for the considered schemes.

Fig. \ref{fig12} shows the BER performance of coherently detected schemes considering FO of \(\Delta f = 0.2\) Hz in the AWGN channel for \(\lambda = 8\). Again, we can observe that the performance of the schemes which transmit information in the IQ components is considerably affected by the FO. It can also be observed that DM-TDM-CSS requires \(1\) dB higher \(E_\mathrm{b}/N_0\) to attain a BER of \(10^{-3}\) while yielding almost double SE. The BER performance of the proposed DM-TDM-CSS is reasonably robust against the FO, making it a more practical alternative to other schemes, particularly IQ-TDM-CSS.

From Fig. \ref{fig13}, which depicts the BER performance of non-coherent detection for the considered approaches, we gather that DM-TDM-CSS performs well relative to the other counterparts. To be more precise, it nearly needs an additional \(0.8\) dB \(E_\mathrm{b}/N_0\) to attain a BER of \(10^{-3}\) relative to DM-CSS, and \(0.9\) dB higher \(E_\mathrm{b}/N_0\) compared to TDM-CSS and IQ-CSS. However,  it is highlighted that in addition to being capable of achieving lower SE, the performance of DM-CSS and IQ-CSS is severely affected by the PO. 
\vspace{-2mm}
\subsection{Advantages and Disadvantages of DM-TDM-CSS}
From the results in the previous section, DM-TDM-CSS has some apparent advantages. Firstly, we observe that it can achieve a higher SE than other alternatives, apart from IQ-TDM-CSS, which transmits only four additional bits per symbol. Secondly, enhanced resilience against the PO and FO compared to other counterparts. It is noticeable that IQ-CSS, IQ-TDM-CSS, and non-coherently detected DM-CSS are very much affected by the PO. The performance of IQ-CSS and IQ-TDM-CSS also degrades due to FO. Thirdly, it manifests a robust performance in the frequency selective channel compared to other counterparts, such as IQ-CSS and IQ-TDM-CSS. Lastly, there is no other CSS approach capable of both coherent and non-coherent detection, and yields a SE of \(\sfrac{(4\lambda-4)}{M}\) at the same time. 

The most significant disadvantage of DM-TDM-CSS is that it does not possess a constant envelop, which could be a concerning point in its practical realization. However, it can also be gathered from the literature that most CSS schemes capable of achieving higher SE than classical LoRa do not possess a constant envelop because these schemes either use a PS or additional FS to attain a higher SE. 
\section{Conclusions}
In this work, we have proposed DM-TDM-CSS as an alternative to the state-of-the-art schemes available in the literature. DM-TDM-CSS has some definite advantages over other counterparts. Besides offering the possibility of coherent and non-coherent detection, it can attain higher maximum achievable SE. Moreover, DM-TDM-CSS is robust against both the PO and the carrier FO. It is shown that the BER performance of the proposed approach is better than schemes offering similar spectral efficiencies in frequency selective fading channels. It is also shown through mathematical analysis that DM-TDM-CSS symbols are not orthogonal. Additionally, the interference caused by the chirp symbols is mathematically evaluated, showing that the simultaneous activation of two chirp symbols having different chirp rates causes interference with each other. It is foreseen that the advantages of DM-TDM-CSS pointed out in this work could encourage further research into the scheme.

\bibliographystyle{unsrt}
\bibliography{biblio}

\appendices
\section{Generalized Quadratic Gauss Sum}\label{apdx}
Any equation like
\begin{equation}\label{eq1_apdx}
I_\pm = \sum_{n=0}^{\vert c\vert -1}\exp\left\{j\frac{\pi}{c}\left(bn\pm an^2\right)\right\}
\end{equation}
takes the form of \textit{generalized quadratic Gauss sum}. Let \(a\), \(b\), and \(c\) be integers with \(ac \neq 0\) and \(ac + b\) even, then one has the following analog of the quadratic reciprocity law for Gauss sum \cite{berndt1998gauss}
\begin{equation}\label{eq2_apdx}
\begin{split}
\sum_{n=0}^{\vert c\vert -1}\exp\left\{j\frac{\pi}{c}\left(bn + an^2\right)\right\} = &\sqrt{\left\vert \frac{c}{a}\right\vert}\exp\left\{j\frac{\pi}{4ac}\left(\vert ac \vert-b^2\right)\right\}\\
&\sum_{n=0}^{\vert a\vert -1}\exp\left\{-j\frac{\pi}{a}\left(bn + cn^2\right)\right\}
\end{split}
\end{equation}
\end{document}